# Structural characterization of the rutile-$GeO_2$/$TiO_2$ (001) heterointerface

Running title: Rutile-$GeO_2$/$TiO_2$ (001) Heterointerface

Running Authors: Rahaman et al.

Imteaz Rahaman[1], Martha R. McCartney[2], David J. Smith[2], and Kai Fu[1,a]

[1]Department of Electrical and Computer Engineering, The University of Utah, Salt Lake City, UT 84112, USA

[2]Department of Physics, Arizona State University, Tempe, AZ 85287, USA

[a)] Electronic mail: kai.Fu@utah.edu

Rutile-$GeO_2$ (r-$GeO_2$) has emerged as a promising ultrawide-bandgap (UWBG) semiconductor for future high-power device applications because of its favorable combination of physical and electrical properties. This study has used high-resolution transmission electron microscopy (HR-TEM) to compare heterointerfaces between thin films of $GeO_2$ grown by metal-organic chemical vapor deposition on closely-lattice-matched rutile $TiO_2$ (r-$TiO_2$) substrates at temperatures of 840° C and 925° C. Cross-sectional TEM observations reveal interfaces with two entirely different structures: an amorphous $GeO_x$/r-$TiO_2$ heterostructure with an abrupt, highly faceted and non-planar interface, or a fully epitaxial r-$GeO_2$/r-$TiO_2$ heterostructure characterized by a more wavy but often facetted interface. Structural analysis of the latter sample revealed a high density of strain-relieving interfacial dislocations, identified primarily as Lomer-type or 60°-type, that accommodated the ~4% lattice mismatch between the two materials, while chemical analysis showed very limited interfacial intermixing. Crystallographic analysis showed that the interface faceting was associated with rutile {101} facets, with characteristic facet angles of ~114°; these observations support thermally driven $TiO_2$

(001) substrate faceting rather than an interface reconstruction resulting from $GeO_2$ epitaxy. Overall, these findings provide useful insights into the mechanisms governing interface formation and strain relaxation in r-$GeO_2$/r-$TiO_2$ heterostructures and suggest that further refinement of growth conditions should enable improved epitaxial growth of r-$GeO_2$ for future ultrawide-bandgap electronic device applications.

# I. INTRODUCTION

Ultrawide bandgap (UWBG) semiconductors, including α- and β-$Ga_2O_3$ ($E_g$= 5.3–5.6 eV and 4.48–4.9 eV, respectively),[1–4] AlGaN ($E_g$ = 3.4–6.2 eV),[5-6] c-BN ($E_g$ = 6.1–6.4 eV),[7] and diamond ($E_g$ = 5.5 eV)[8] have attracted much recent attention as promising materials for next-generation power electronics, due to their large band gaps, high breakdown strengths, and stability under extreme operating conditions.[8–12] However, each of these materials has practical constraints, ranging from doping and thermal management to processing complexity and substrate scalability, that often limit their implementation in devices.[13–16] These intrinsic challenges underscore the importance of broadening the UWBG materials landscape to identify alternative platforms that would offer similar and/or complementary properties.

Rutile germanium dioxide (r-$GeO_2$) has recently emerged as a promising UWBG semiconductor due to its unique combination of electronic, thermal, and structural properties.[17] r-$GeO_2$ has a band gap of ~4.4–5.1 eV,[17–20] high electron mobilities of 244 $cm^2/V{\cdot}s$ ($\perp c$) and 377 $cm^2/V{\cdot}s$ ($\parallel c$),[21] and an outstanding *n*-type Baliga figure of merit of 27,000–35,000 × $10^6$ $V^2\ \Omega^{-1}\ cm^{-2}$.[17] The thermal conductivity — 37 $W{\cdot}m^{-1}{\cdot}K^{-1}$ along the *a*-axis, 58 $W{\cdot}m^{-1}{\cdot}K^{-1}$ along the *c*-axis, and ~51 $W{\cdot}m^{-1}{\cdot}K^{-1}$ on average — exceeds that of $Ga_2O_3$ by more than a factor of two.[22] Both *n*- and *p*-type conductivity can be supported due to low dopant ionization energies,[18,23] with predicted hole mobilities of 27 $cm^2/V \cdot s$ ($\perp c$) and 29 $cm^2/V \cdot s$ ($\parallel c$).[18,21,23] Unlike the hexagonal (quartz) polymorph,[24] r-$GeO_2$ exhibits low solubility in humid environments, enhancing stability. Structurally, its crystallization in the rutile lattice with octahedrally coordinated Ge atoms imparts high density, high refractive index, and strong mechanical hardness.[25] The band gap can also be tuned by alloying with other rutile-type oxides such as $SnO_2$, $TiO_2$, and $SiO_2$,[26,27] analogous to III–V bandgap engineering.[28] These attributes combine to make r-$GeO_2$ a compelling prospect for high-power electronic devices as well as emerging applications in optoelectronics, photocatalysis, and gas sensing.[29,30]

Enabling the growth of high-quality r-$GeO_2$ films is critical to realizing the full potential of r-GeO2 in electronic applications. The growth of r-$GeO_2$ thin films has been demonstrated on r-$TiO_2$ and sapphire substrates using molecular beam epitaxy,[31] mist-chemical vapor deposition,[32,33] pulsed laser deposition,[19,34,35] and metal-organic chemical vapor deposition (MOCVD).[36–38] Our recent work has revealed a distinctive square-patterned faceted surface morphology in MOCVD-grown films.[37,39] The r-$TiO_2$(001) surface is known to have high surface energy and to undergo thermally driven faceting.[40,41] Related faceted morphologies have also been reported in $VO_2/TiO_2$(001) heterostructures.[42] However, achieving phase-pure rutile-$GeO_2$ remains a significant challenge due to the strong competition between the amorphous, quartz, and rutile polymorphs.[36,43] Published work on effective phase control is still very limited.[39] The film–substrate interface is recognized as pivotal in governing phase selection, strain relaxation, defect formation, and overall crystalline quality, yet it has received little direct attention. A more detailed knowledge of this interface could thus be highly beneficial for gaining insights into the stabilization of the rutile phase. In this study, we report high-resolution structural and chemical characterization of interfaces between $GeO_2$ films grown by MOCVD on rutile (001) $TiO_2$ substrates.

## II. EXPERIMENTAL

The undoped $GeO_2$ thin films were deposited using an Agilis MOCVD system (Agnitron Technology) on r-$TiO_2$ (001) substrates, chosen for their favorable compatibility (lattice misfit ~4%), thereby reducing strain with $GeO_2$.[29,32] Tetraethylgermane (TEGe) was used as the germanium precursor, with high-purity $O_2$ supplied as the oxidant, and argon was used both as the carrier gas and the shroud gas. The precursor flow rates were maintained at $8.94 \times 10^{-2}$ mol/min for $O_2$ and $1.35 \times 10^{-5}$ mol/min for TEGe, while the susceptor was rotated at 300 RPM throughout the deposition process to promote uniform film growth. Prior to loading into the MOCVD chamber, the substrates were cleaned with a Piranha solution (3:1 $H_2SO_4$:$H_2O_2$), then sequentially rinsed in acetone, isopropanol, and deionized water. The two samples described here were grown at substrate temperatures of 840 °C and 925 °C under chamber pressures of 80 Torr; the corresponding samples are referred to in the following as T-23 and T-25, respectively. [37] These two growth temperatures provide an interesting contrast in the observed interface structures.

## III. RESULTS AND DISCUSSION

After completion of growth, samples of T-23 and T-25 were prepared for observation by transmission electron microscopy (TEM) techniques in the cross-section geometry so that the (001) substrate surface normal would be perpendicular to the direction of the incident electron beam. Thinning to electron transparency was performed using Ga ion-beam milling with a Thermo Fisher Helios 5UX FIB-SEM dual-beam instrument, initially at 30 keV, followed by milling at 5 keV and 2 keV to minimize surface damage from the ion-milling process. An image-corrected FEI Titan 80-300 operated at 300 kV and a probe-corrected JEOL ARM-200F operated at 200 keV were used for high-resolution imaging, and a Thermo Fisher Talos 200i operated at 200 kV was used for microanalysis with energy-dispersive X-ray spectroscopy (EDXS).

Figure 1 compares the microstructural characteristics of the two $GeO_x/TiO_2$ heterostructures. Figure 1(a) shows a cross-sectional TEM micrograph of Sample T-23, which was grown at a substrate temperature of 840°C. The enlarged image in Fig. 1(b) reveals a highly faceted interfacial morphology between the deposited $GeO_x$ layer and the r-$TiO_2$ (001) substrate, indicating that substantial restructuring of the substrate surface has taken place during growth. The higher-magnification TEM image in Fig. 1(c) further demonstrates that the upper surface of the $TiO_2$ substrate has become significantly faceted. In contrast to the crystalline substrate, the deposited overlayer shows no discernible lattice fringes, indicating an amorphous structure. The inset FFT, obtained partly from the $GeO_x$ region, exhibits diffuse halo features, confirming the amorphous nature of the deposited $GeO_x$ film.

Figures 1(d–f) present equivalent cross-sectional images of Sample T-25, grown at 925°C, which exhibits a markedly different microstructure. The low-magnification image in Fig. 1(d) shows the presence of threading defects in the continuous r-$GeO_2$ layer, which is observed to be fully crystalline without any misoriented domains, while the inset selected-area electron diffraction (SAED) pattern indicates that the r-$GeO_2$ layer is epitaxially aligned with the r-$TiO_2$ (001)-oriented substrate. The enlargement in Fig. 1(e) shows that the r-$GeO_2$/r-$TiO_2$ heterointerface has developed a somewhat wavy morphology, which is unlike the highly faceted appearance of the interface of Sample T-23 shown in Fig. 1(b). The higher-magnification image in Fig. 1(f) suggests the presence of significant interfacial strain, consistent with the anticipated high density of misfit dislocations resulting from the lattice mismatch (~4%) between the two materials. Analysis of the two sets of closely aligned spots visible in the inset FFT establishes the epitaxial relationship of (001)r-$GeO_2$||(001) r-$TiO_2$ and [100] r-$GeO_2$|| [100] r-$TiO_2$.

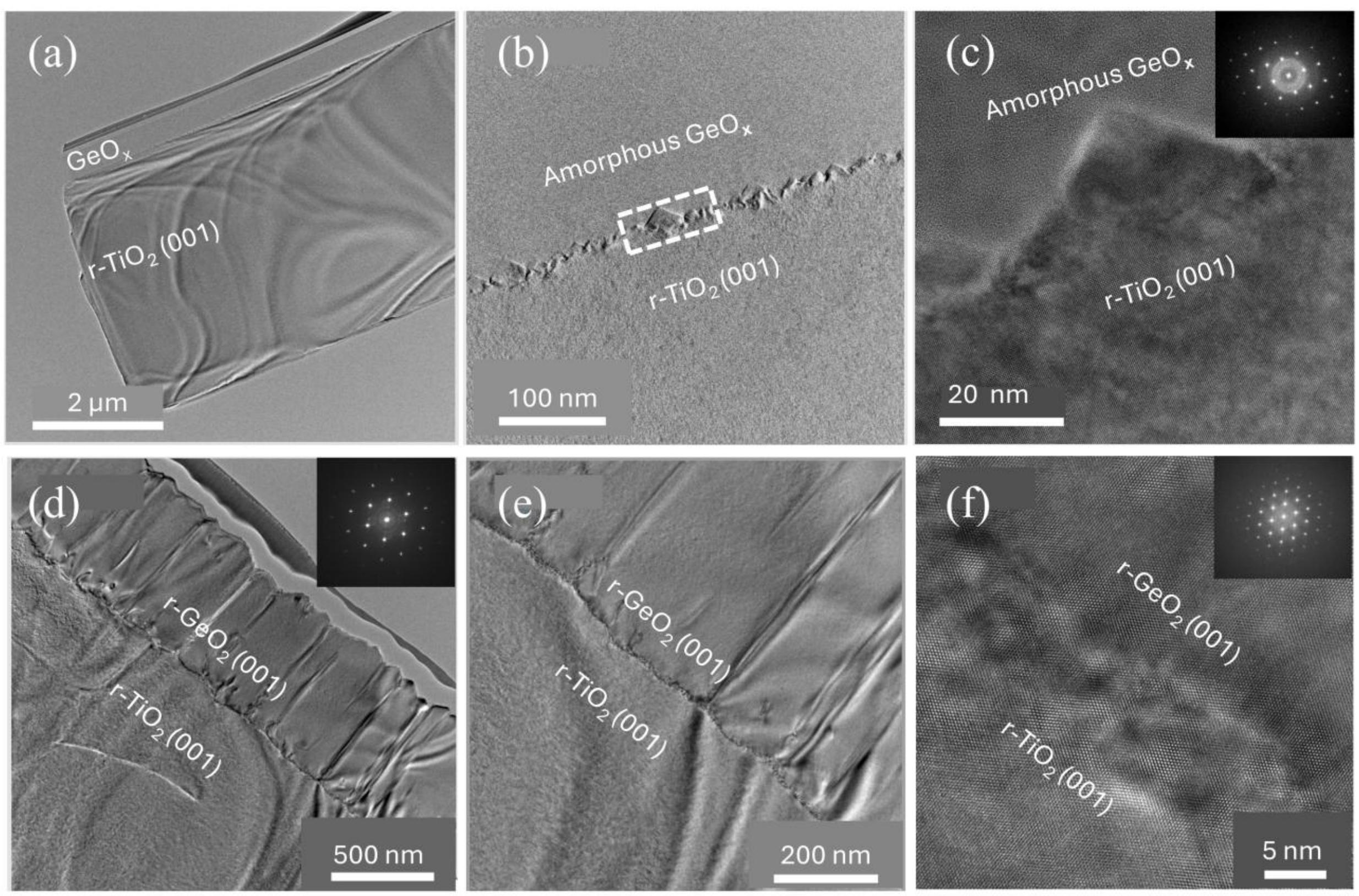

FIG. 1. Cross-sectional TEM characterization of amorphous $GeO_x$/r-$TiO_2$ and epitaxial r-$GeO_2$/r-$TiO_2$ heterostructures. (a–c) Amorphous $GeO_x$/r-$TiO_2$ heterostructure: (a) low-magnification TEM image, (b) enlarged view of the faceted $GeO_x$/r-$TiO_2$ interface, and

(c) high-resolution TEM image showing an amorphous $GeO_x$ overlayer. The inset FFT exhibits diffuse halo features characteristic of an amorphous structure. (d–f) Epitaxial r-$GeO_2$/r-$TiO_2$ heterostructure: (d) low-magnification TEM image showing a crystalline r-$GeO_2$ layer and threading defects. The inset SAED pattern confirms the epitaxial relationship between the r-$GeO_2$ film and the r-$TiO_2$ substrate. (e) Enlarged view of the heterointerface exhibiting wavy morphology. (f) High-resolution TEM image of the r-$GeO_2$/r-$TiO_2$ interface. The inset FFT exhibits two closely aligned diffraction patterns confirming epitaxial growth of r-$GeO_2$ on r-$TiO_2$.

Elemental mapping using EDXS was used to assess the chemical separation between the $GeO_2$ layer and the $TiO_2$ substrate in both samples. Figure 2 shows the EDXS analysis for an interface region of the epitaxial r-$GeO_2$/r-$TiO_2$ (001) heterostructure (Sample T-25). The distinct contrast between the two layers, apparent in the bright-field STEM image in Fig. 2(a), again emphasizes the non-planar, wavy nature of the r-$GeO_2$/r-$TiO_2$ interface. Figure 2(b) is the corresponding dark-field STEM image, while Fig. 2(c) shows separate elemental maps for O, Ti, and Ge. Oxygen is distributed uniformly throughout both layers, while no extended regions of significant Ge diffusion into the $TiO_2$ substrate or Ti diffusion into the $GeO_2$ film are observed within the spatial resolution of the measurement. These latter two elemental maps emphasize that the r-$GeO_2$ epilayer and the r-$TiO_2$ substrate are chemically separated following high-temperature growth, again confirming the formation of a well-defined epitaxial oxide-oxide heterointerface.

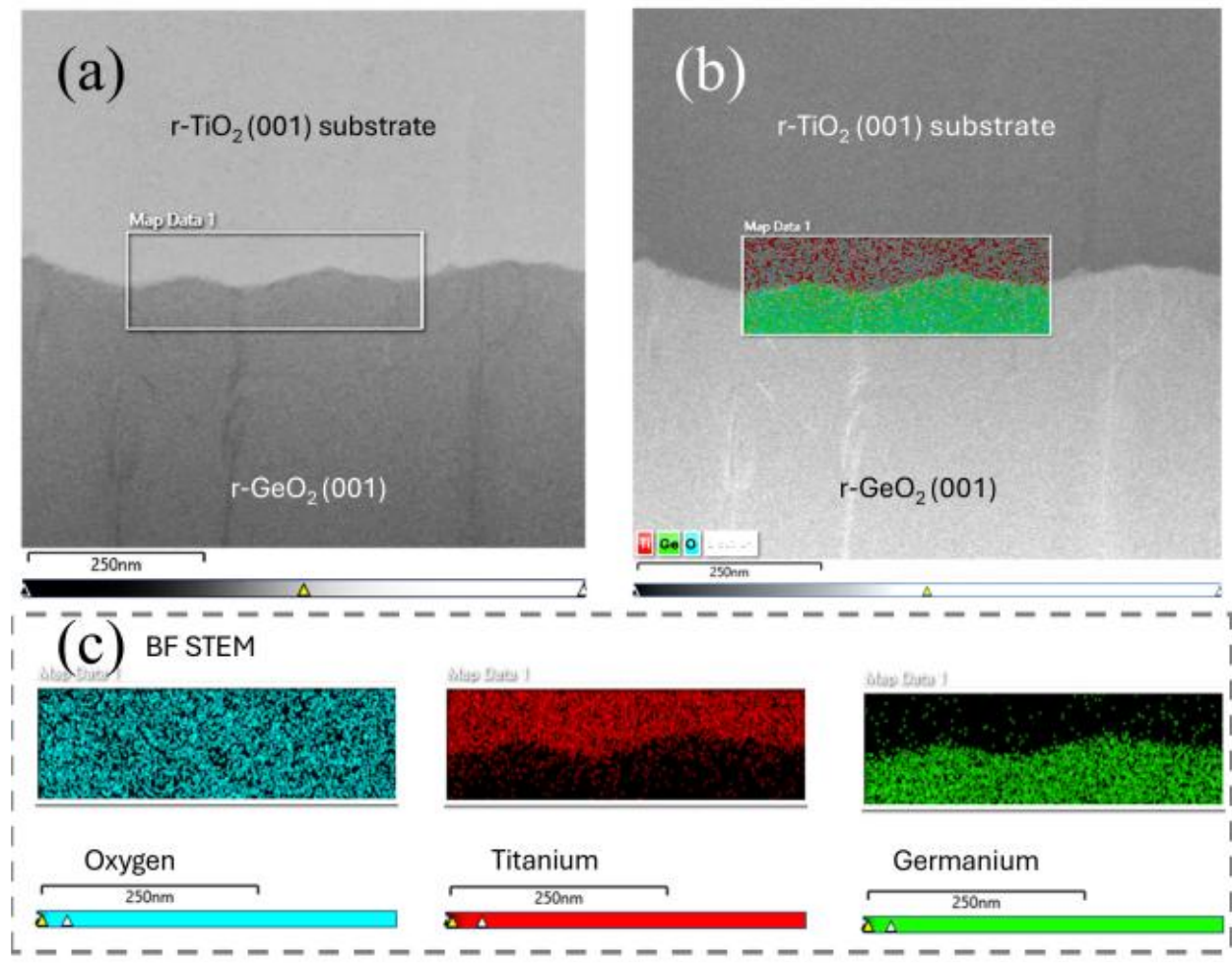

FIG. 2. EDXS elemental mapping of r-$GeO_2$/r-$TiO_2$ (001) heterostructure. (a) Bright-field STEM image highlighting the region selected for mapping. (b) Dark-field STEM image with overlay of elemental maps showing the distribution of Ge, Ti, and O. (c) Individual EDXS maps of Oxygen, Titanium, and Germanium, confirming sharp chemical separation of Ti and Ge between the r-$GeO_2$ film and the r-$TiO_2$ (001) substrate.

Atomic-resolution TEM imaging was used to provide enhanced details of the epitaxial r-$GeO_2$/r-$TiO_2$ interface in Sample T-25. Figure 3(a) is a representative example that confirms direct crystalline contact between the r-$GeO_2$ film and the r-$TiO_2$ substrate. The interface remains structurally coherent over many-nanometer scales, and there is no evidence of an amorphous interlayer between the two materials. Burgers circuit analysis of defects decorating the highly strained interface indicated that the most common types of interfacial misfit dislocations were Lomer edge-type and 60°-type dislocations, albeit with slightly different angles for the latter defects reflecting the tetragonal crystal structure of rutile. More complex dislocation networks were also visible, as illustrated in the interface region of Fig. 3(a) and the corresponding Fourier-filtered image in Fig. 3(b), which enhances the visibility of lattice discontinuities and defect cores along the heterointerface.

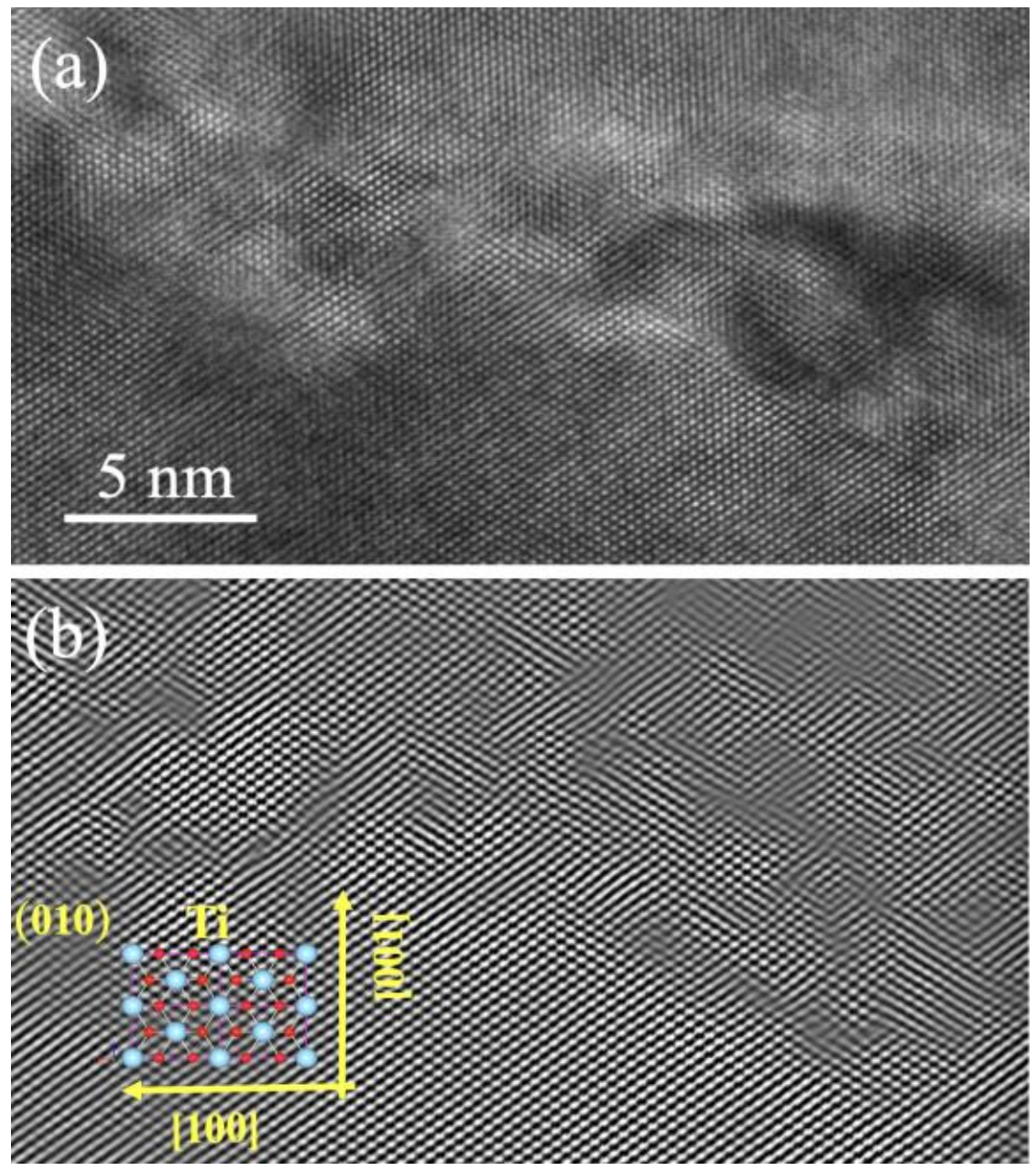

FIG. 3. (a) Atomic-resolution image of the r-$GeO_2$/r-$TiO_2$ (001) interface showing a complex array of interfacial misfit dislocations; (b) Fourier-filtered image showing lattice discontinuities corresponding to dislocation terminations. The bottom inset shows the corresponding $TiO_2$ structural model.

Because of its enhanced sensitivity to elemental atomic number, high-angle annular dark-field (HAADF) imaging (i.e., “Z-contrast”) provides a powerful approach for elucidating details of the structural (and chemical) profile of the r-$GeO_2$/r-$TiO_2$ interface region. Figure 4(a) shows a representative HAADF image of the interface, where the crystalline r-$GeO_2$ film follows the faceted topography of the underlying $TiO_2$ substrate. There is again no evidence for the presence of any amorphous region immediately above the substrate surface. The enlarged images shown in Figs. 4(b, c) highlight misfit dislocations at the r-$GeO_2$/r-$TiO_2$ boundary, where the arrows indicate locations of the dislocation cores. These defects act as strain-relief centers that partially accommodate the lattice mismatch between the two materials. The surrounding lattice remains well-ordered, indicating that epitaxial registry is maintained across the interface while misfit strain is relieved through localized defect formation.

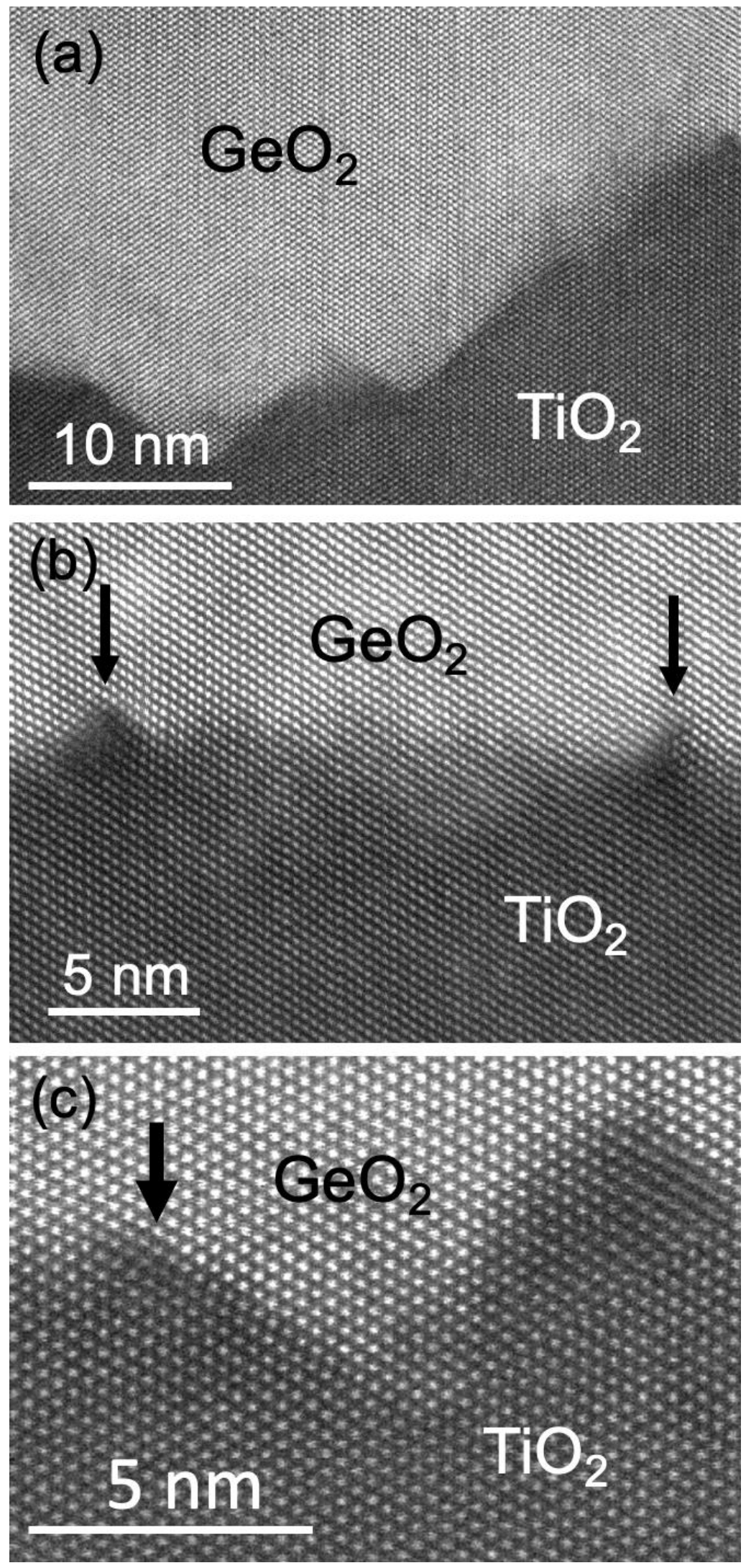

FIG. 4. HAADF imaging of the epitaxial r-$GeO_2$/r-$TiO_2$ (001) interface in Sample T-25. (a) Region of uneven and faceted r-$GeO_2$/r-$TiO_2$ interface. (b–c) Enlarged atomic-resolution images highlighting representative interfacial dislocations with extra half-planes (arrowed) that accommodate r-$GeO_2$/r-$TiO_2$ lattice mismatch while preserving overall epitaxial registry.

HAADF) imaging (i.e., “Z-contrast”) was also used to elucidate further details of the structural profile of the r-$GeO_2$/r-$TiO_2$ interface region. A representative HAADF image is shown in Fig. 5(a); the measured facet angles are approximately 114°. The repeated occurrence of this distinctive microstructure at multiple locations along the interface indicates that this type of faceting is crystallographically driven rather than arising from random surface roughening. The faceted interface is further examined in the enlarged atomic-resolution HAADF image shown in Fig. 5(b), which, for reference, also includes an outline of the $TiO_2$ unit cell and the corresponding crystallographic directions. The measured facet angle closely matches the angle formed by adjacent {101} planes in the rutile crystal structure, while the facet boundaries align with the [101] crystallographic direction. The dashed guidelines emphasize the geometric relationship between the interface facets and the underlying rutile lattice, whereas the substrate surface remains oriented along the (001) plane. The excellent agreement between the experimentally observed facet geometry and the expected rutile crystallography indicates that the faceted interface morphology is strongly governed by the crystal symmetry of the r-$GeO_2$/$TiO_2$ heterosystem and most likely originates from preferential stabilization of rutile {101} facets during growth.

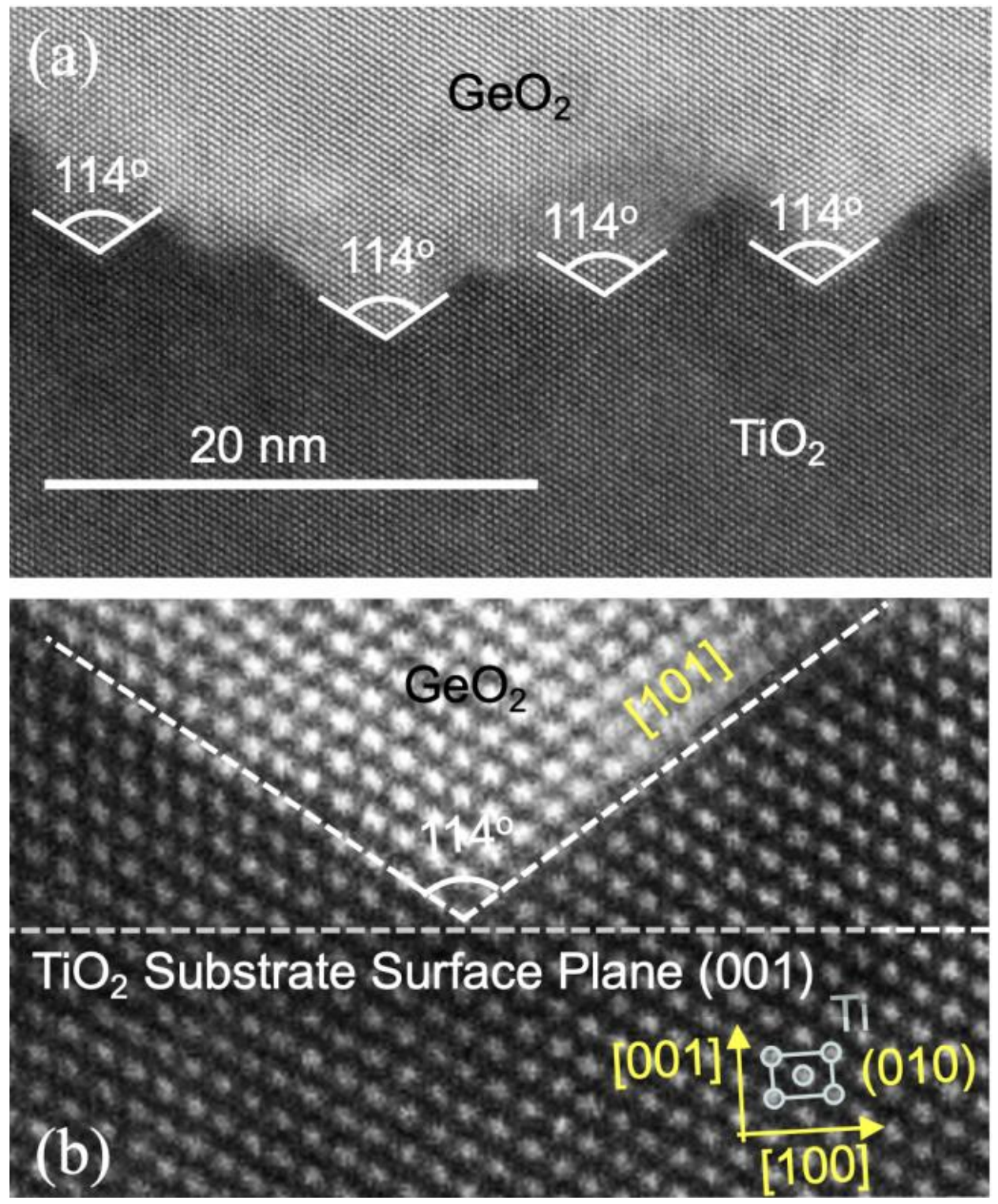

FIG. 5. Crystallographic analysis of the faceted region of the r-$GeO_2$/r-$TiO_2$ (001) heterointerface. (a) HAADF image showing the faceted interface morphology. Measured facet angles are ~ 114°, consistent with the characteristic angle between adjacent rutile {101} planes. (b) Enlarged image highlighting facet geometry and crystallographic orientation relationship. Dashed lines indicate facet boundaries. The inset shows the rutile $TiO_2$ crystal structure and corresponding crystallographic directions used for indexing.

# IV. SUMMARY AND CONCLUSIONS

In summary, high-resolution TEM characterization has revealed two distinct interfacial structures in $GeO_2$ films grown on r-$TiO_2$ (001) substrates at different MOCVD growth temperatures. Growth at 840°C yielded a highly faceted, non-planar interface with an amorphous $GeO_x$ overlayer. In contrast, growth at 925°C produced a fully crystalline r-$GeO_2$ layer but with a more wavy, often faceted interfacial morphology. Atomic-resolution imaging confirmed the absence of any amorphous interfacial layer in the high-temperature sample and also revealed a significant density of strain-relieving interfacial defects, including Lomer edge-type and 60°-type dislocations, which were attributed to the ~4% lattice mismatch between r-$GeO_2$ and r-$TiO_2$. Chemical analysis, as well as HAADF-STEM imaging, demonstrated chemically distinct, yet locally non-uniform, compositional transitions across the interface. The occurrence of crystallographic faceting beneath both amorphous and crystalline overlayers supports high-temperature evolution of r-$TiO_2$(001) as the origin of the interface topography. These results confirm the strong correlation between growth temperature, interfacial morphology, strain relaxation, and crystalline phase formation in MOCVD-grown r-$GeO_2$ heterostructures. The present findings demonstrate the potential to stabilize the growth of heteroepitaxial rutile-$GeO_2$, thereby advancing the development and optimization of r-$GeO_2$-based UWBG electronic materials and devices.

# ACKNOWLEDGMENTS

The authors acknowledge support from the University of Utah start-up fund, and use of the University of Utah Nanofab shared facilities, which are supported in part by the NSF MRSEC Program under Award DMR-112125. The authors also acknowledge the use of facilities within the John M. Cowley Center for High Resolution Electron Microscopy at Arizona State University, supported in part by NSF program NNCI-ECCS-1542160, and thank Dr. Piyush Haluai for assistance with recording EDXS data.

# AUTHOR DECLARATIONS

**Conflicts of Interest**

The authors have no conflicts to disclose.

**Ethics Approval**

Ethics approval was not required because this study did not involve human participants or animals.

**Author Contributions**

**Imteaz Rahaman**: Methodology (Lead), Formal analysis (Supporting); Investigation (Supporting); Writing-original draft (Lead). **Martha R. McCartney**: Data curation (Equal); Formal analysis (Equal), Resources (Equal), Supervision (Equal), Project administration (Equal). **David J. Smith**: Conceptualization (lead), Data curation (Equal); Formal analysis (Equal), Resources (Equal), Supervision (Equal), Project administration (Equal), Writing – review & editing (Equal). **Kai Fu**: Supervision (Equal), Project administration (Equal), Resources (Equal), Conceptualization (Equal), Writing – review & editing (Equal).

# DATA AVAILABILITY

The data that support the findings of this study are available from the corresponding authors upon reasonable request.